\documentclass[12pt,a4paper]{article}
\usepackage{mathrsfs}
\usepackage{epsfig}
\begin{document}
\title{Lepton flavor violating signals of a little Higgs model at the high energy linear $e^{+}e^{-}$ colliders}
\author{Chong-Xing Yue and Shuang Zhao\\
{\small Department of Physics, Liaoning Normal University, Dalian
116029, China
\thanks{E-mail:cxyue@lnnu.edu.cn}}\\}
\date{\today}
\maketitle
\begin{abstract}
\hspace{5mm}Littlest Higgs $(LH)$ model predicts the existence of
the doubly charged scalars $\Phi^{\pm\pm}$, which generally have
large flavor changing couplings to leptons. We calculate the
contributions of $\Phi^{\pm\pm}$ to the lepton flavor violating
$(LFV)$ processes $l_{i}\rightarrow l_{j}\gamma$ and
$l_{i}\rightarrow l_{j}l_{k}l_{k}$, and compare our numerical
results with the current experimental upper limits on these
processes. We find that some of these processes can give severe
constraints on the coupling constant $Y_{ij}$ and the mass parameter
$M_{\Phi}$. Taking into account the constraints on these free
parameters, we further discuss the possible lepton flavor violating
signals of $\Phi^{\pm\pm}$ at the high energy linear $e^{+}e^{-}$
collider $(ILC)$ experiments. Our numerical results show that the
possible signals of $\Phi^{\pm\pm}$ might be detected via the
subprocesses $e^{\pm}e^{\pm}\rightarrow l^{\pm}l^{\pm}$ in the
future $ILC$ experiments.

\end {abstract}

\vspace{0.8cm}

\newpage
\section*{I. Introduction}

\hspace{5mm}It is well known that the individual lepton numbers
$L_{e}, L_{\mu}$, and $L_{\tau}$ are automatically conserved and the
tree level lepton flavor violating $(LFV)$ processes are absent in
the standard model $(SM)$. However, the neutrino oscillation
experiments have made one believe that neutrinos are massive,
oscillate in flavors, which presently provide the only experimental
hints of new physics and imply that the separated lepton numbers are
not conserved[1]. Thus, the $SM$ requires some modification to
account for the pattern of neutrino mixing, in which the $LFV$
processes are allowed. The observation of the $LFV$ signals in
present or future high energy experiments would be a clear signature
of new physics beyond the $SM$.

Some of popular specific models beyond the $SM$ generally predict
 the presence of new particles, such as new gauge bosons and new
 scalars, which can naturally lead to the tree level $LFV$ coupling.
 In general, these new particles could enhance branching ratios for
 some $LFV$ processes and perhaps bring them into the observable
 threshold of the present and next generations of collider
 experiments. Furthermore, nonobservability of these $LFV$ processes
 can lead to strong constraints on the free parameters of new
 physics. Thus, studying the possible $LFV$ signals of new physics
 in various high energy collider experiments is very interesting and
 needed.

 Little Higgs models[2] employ an extended set of global and gauge
 symmetries in order to avoid the one-loop quadratic divergences and
 thus provide a new method to solve the hierarchy between the $TeV$
 scale of possible new physics and the electroweak scale $\nu=246 GeV=(\sqrt{2}
 G_{F})^{-\frac{1}{2}}$. In this kind of models, the Higgs boson is a
 pseudo-Goldstone boson of a global symmetry which is spontaneously
 broken at some high scales. Electroweak symmetry breaking $(EWSB)$
 is induced by radiative corrections leading to a Coleman-Weinberg
 type of potential. Quadratic divergence cancellation of radiative
 corrections to the Higgs boson mass are due to contributions from
 new particles with the same spin as the $SM$ particles. This type
 of models can be regarded as one of the important candidates of the
 new physics beyond the $SM$.

 The littlest Higgs model $(LH)$[3] is one of the simplest and
 phenomenologically viable models, which realizes the little Higgs
 idea. Recently, using of the fact that the $LH$ model contains a
 complex triplet Higgs boson $\Phi$, Refs.[4,5,6] have discussed the
 possibility to introduce lepton number violating interactions and generation
 of neutrino mass in the little Higgs scenario. Ref.[5] has shown
 that most satisfactory way of incorporating neutrino masses is to
 include a lepton number violating interaction between the triplet
 scalars and lepton doublets. The tree level neutrino masses are
 mainly generated by the vacuum expectation value $(VEV)$ $\nu'$ of
 the complex triplet $\Phi$, which does not affect the cancellation
 of quadratic divergences in the Higgs mass. The neutrino masses can
 be given by the term $Y_{ij}\nu'$, in which $Y_{ij}$ ($i, j$ are
 generation indices) is the Yukawa coupling constant. As long as the
 triplet $VEV$ $\nu'$ is restricted to be extremely small, the value
 of $Y_{ij}$ is of natural order one, i.e. $Y_{ij}$ $\approx$ 1, which
 might produce large contributions to some of $LFV$ processes[6,7].

 The aim of this paper is to study the contributions of the $LFV$
 couplings predicted by the $LH$ model to the $LFV$ processes $l_{i} \rightarrow l_{j}
 \gamma$ and $l_{i} \rightarrow l_{j}l_{k}l_{k}$ and compare our
 numerical results with the present experimental bounds on these
 $LFV$ processes, and see whether the constraints on the free
 parameter $Y_{ij}$ can be obtained. We further calculate
 the contributions of the $LH$ model to the $LFV$ processes $e^{\pm}e^{\pm}\rightarrow l_{i}^{\pm}l_{j}^{\pm}$
 and $e^{+}e^{-} \rightarrow l_{i}^{\pm}l_{j}^{\pm}$($l_{i}$ or $l_{j}$$\neq e$) ,
 and discuss the possibility of detecting the $LFV$
 signals of the $LH$ model via these processes in the future
 high energy linear $e^{+}e^{-}$ collider $(ILC)$ experiments.

 This paper is organized as follows. Section II contains a short
 summary of the relevant $LFV$ couplings of the scalars (doubly
 charged scalar $\Phi^{\pm\pm}$, charged scalars $\Phi^{\pm}$, and
 neutral scalar $\Phi^{0}$) to lepton doublets. The contributions of
 these $LFV$ couplings to the $LFV$ processes $l_{i} \rightarrow
 l_{j}\gamma$ and $l_{i} \rightarrow l_{j}l_{k}l_{k}$ are calculated
 in section III. Using the current experimental upper limits on these $LFV$
 processes, we try to give the constraints on the coupling constant
 $Y_{ij}$ in this section. Section IV is devoted to the computation
 of the production cross sections of the $LFV$ processes $e^{\pm}e^{\pm}\rightarrow l_{i}^{\pm}l_{j}^{\pm}$
 and $e^{+}e^{-} \rightarrow l_{i}^{\pm}l_{j}^{\pm}$ induced by the doubly charged scalars
 $\Phi^{\pm\pm}$. Some phenomenological analyses are also included in this section. Our
 conclusions are given in section V.

\section*{II. The $LFV$ couplings of the triplet scalars }

\hspace{5mm} The $LH$ model[3] consists of a nonlinear $\sigma$
model with a global $SU(5)$ symmetry and a locally gauged symmetry
$[SU(2) \times U(1)]^{2}$. The global $SU(5)$ symmetry is broken
down to its subgroup $SO(5)$ at a scale $f \sim TeV$, which results
in 14 Goldstone bosons $(GB's)$. Four of these $GB's$ are eaten by
the gauge bosons $(W^{\pm}_{H}, Z_{H}, B_{H})$, resulting from the
breaking of $[SU(2) \times U(1)]^{2}$, giving them masses. The Higgs
boson remains as a light pseudo Goldstone boson and other $GB's$
give masses to the $SM$ gauge bosons and form a scalar triplet
$\Phi$. The complex triplet $\Phi$ offers a chance to introduce
lepton number violating interactions in the theory.

In the context of the $LH$ model, the lepton number violating
interaction which is invariant under the full gauge group, can be
written as[5,7]:
\begin{equation}
\mathscr{L}=-\frac{1}{2}Y_{ij}(L_{i}^{T})_{\alpha}\Sigma
^{\ast}_{\alpha\beta}C^{-1}(L_{j}^{T})_{\beta}+h.c.
\end{equation}
Where $i$ and $j$ are generation indices, $\alpha$ and $\beta$ (= 1,
2) are $SU(5)$ indices, and $L^{T}=(l_{L},\nu_{L})$ is a left handed
lepton doublet. $Y_{ij}$ is the Yukawa coupling constant and $C$ is
the charge-conjugation operator. Because of non-linear nature of
$\Sigma_{\alpha\beta}^{\ast}$, this interaction can give rise to a
mass matrix for the neutrinos as:
\begin{equation}
M_{ij}=Y_{ij}(\nu'+\frac{\nu^{2}}{4f}).
\end{equation}

 One can see from Eq.(2) that, if we would like to stabilize the
Higgs mass and at the same time ensure neutrino masses consistent
with experimental data[8], the coupling constant $Y_{ij}$ must be of
order $10^{-11}$, which is unnaturally small. However, it has been
shown[4,5] that the lepton number violating interaction only
involving the complex scalar triplet $\Phi$ can give a neutrino mass
matrix $M_{ij}=Y_{ij}\nu'$. Considering the current bounds on the
neutrino mass[8], there should be:
\begin{equation}
Y_{ij}\nu'\sim10^{-10}GeV.
\end{equation}
Thus, the coupling constant $Y_{ij}$ can naturally be of order one
or at least not be unnaturally small provided the $VEV$ $\nu'$ of
the triplet scalar $\Phi$ is restricted to be extremely small.

In this scenario, the triplet scalar $\Phi$ has the $LFV$ couplings
to the left handed lepton pairs, which can be written as[5]:
\begin{equation}
\mathscr{L}_{LFV}=Y_{ij}[l_{Li}^{T}C^{-1}l_{Lj}\Phi^{++}+\frac{1}{\sqrt{2}}
(\nu_{Li}^{T}C^{-1}l_{Lj}+l_{Li}^{T}C^{-1}\nu_{Lj})\Phi^{+}
+\nu_{Li}^{T}C^{-1}\nu_{Lj}\Phi^{0}]+h.c.
\end{equation}
Considering these $LFV$ couplings, Ref.[5] has investigated the
decays of the scalars $\Phi^{\pm\pm}$ and $\Phi^{\pm}$, and found
that the most striking signature comes from the doubly charged
scalars $\Phi^{\pm\pm}$. The constraints on the coupling constant
$Y_{ij}$ and the triplet scalar mass parameter $M_{\Phi}$ coming
from the muon anomalous magnetic moment $a_{\mu}$ and the $LFV$
process $\mu^{-} \rightarrow e^{+}e^{-}e^{-}$ are studied in
Ref.[7]. In the next section, we will calculate the contributions of
the charged scalars $\Phi^{\pm\pm}$ and $\Phi^{\pm}$ to the $LFV$
processes $l_{i} \rightarrow l_{j}\gamma$ and $l_{i} \rightarrow
l_{j}l_{k}l_{k}$.

\section*{III. The charged scalars and the $LFV$ processes $l_{i} \rightarrow
 l_{j}\gamma$  \hspace*{1.0cm} and $l_{i} \rightarrow l_{j}l_{k}l_{k}$ }

 \begin{center}
{
\begin{small}
\begin{tabular}{|c|c|c|}
\hline Decay\ Process&Current\ limit&Bound$(GeV^{-4})$ \\
\hline
$\mu\rightarrow e\gamma$&$1.2\times 10^{-11}$ [10]&----- \\
$\tau\rightarrow e\gamma$&$1.1\times 10^{-7}$ [12]&-----\\
$\tau\rightarrow \mu\gamma$&$6.8\times 10^{-8}$ [13]&-----\\
$\mu\rightarrow 3e$&$1.0\times10^{-12}$ [11]&$\mid Y_{\mu e}Y_{ee}^{\ast}\mid^{2}/M_{\Phi}^{4}
\leq 2.2\times 10^{-19}$\\
$\tau\rightarrow 3e$&$2.0\times 10^{-7}$ [14]&$\mid Y_{\tau e}Y_{ee}^{\ast}\mid^{2}/M_{\Phi}^{4}
\leq 2.4\times 10^{-13}$\\
$\tau\rightarrow 2\mu e$&$3.3\times 10^{-7}$ [14]&$\mid Y_{\tau e}Y_{\mu\mu}^{\ast}\mid^{2}/M_{\Phi}^{4}
\leq 8.1\times 10^{-13}$\\
$\tau\rightarrow 2e\mu$&$2.7\times 10^{-7}$ [14]&$\mid Y_{\tau\mu}Y_{ee}^{\ast}\mid^{2}/M_{\Phi}^{4}
\leq 6.6\times 10^{-13}$\\
$\tau\rightarrow 3\mu$&$1.9\times 10^{-7} [14]$&$\mid Y_{\tau\mu}Y_{\mu\mu}^{\ast}\mid^{2}/M_{\Phi}^{4}
\leq 2.3\times 10^{-13}$\\
\hline\end{tabular}
\end{small} }\end{center}
\hspace{0.3cm} Table 1: The current experimental upper limits on the
branching ratios of some $LFV$ \hspace*{2cm} processes and the
corresponding upper constraints on the free parameters.
\hspace*{1.9cm} \vspace*{0.0cm}

The observation of neutrino oscillations[1] implies that the
individual lepton numbers $L_{e,\mu,\tau}$ are violated, suggesting
the appearance of the $LFV$ processes, such as $l_{i} \rightarrow
l_{j}\gamma$ and $l_{i} \rightarrow l_{j}l_{k}l_{k}$. The branching
ratios of these $LFV$ processes are extremely small in the $SM$ with
right handed neutrinos. For example, Ref.[9] has shown $Br(\mu
\rightarrow e\gamma) <10^{-47}$. Such small branching ratio is
unobservable.

The present experimental upper limits on the branching ratios
$Br(\mu \rightarrow e\gamma)$[10], $Br(\mu \rightarrow 3e)$[11],
$Br(\tau \rightarrow e\gamma)$[12], $Br(\tau \rightarrow
\mu\gamma)$[13], and $Br(\tau \rightarrow l_{i}l_{k}l_{k})$[14] are
given in Table 1. Future experiments with increased sensitivity can
reduce these current limits by a few orders of magnitude(see,
e.g.[15]). In this  section, we will use these data to give the
constraints on the free parameters $Y_{ij}$ and $M_{\Phi}$.

\begin{figure}[htb] \vspace{-7.0cm}
\begin{center}
\epsfig{file=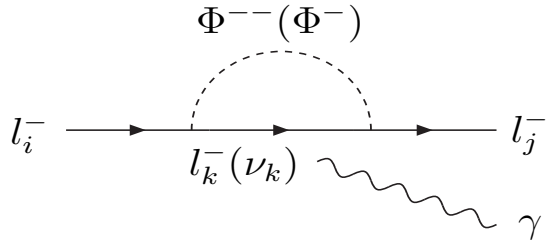,width=800pt,height=1000pt} \vspace{-25.0cm}
 \caption{Feynman diagrams contributing to the radiative decay $l_{i}^{-}\rightarrow l_{j}^{-}\gamma$
  due to the \hspace*{1.8cm}charged scalars $\Phi^{--}(\Phi^{-})$.}
\label{ee} \vspace{-0.5cm}
\end{center}
\end{figure}

The $LFV$ couplings of the charged scalars $\Phi^{--}$ and
$\Phi^{-}$ given in Eq.(4) can lead to the $LFV$ radiative decays
$l_{i}^{-}\rightarrow l_{j}^{-}\gamma$ at the one loop level
mediated by the exchange of the charged scalars $\Phi^{--}$ and
$\Phi^{-}$, as shown in Fig.1. For the doubly charged scalar
$\Phi^{--}$, the photon can be attached either to the internal
lepton line or to the scalar line. For the charged scalar
$\Phi^{-}$, the photon can be only attached to the scalar line[16].

Using Eq.(4), the expression of the branching ratio
$Br(l_{i}^{-}\rightarrow l_{j}^{-}\gamma)$ can be written as at
leading order:
\begin{equation}
Br(l_{i}^{-}\rightarrow l_{j}^{-}\gamma)=\frac{\alpha_{e}}{96\pi
G_{F}^{2}}\sum_{k=\tau,\mu,e}(Y_{ik}Y_{kj}^{\ast})^{2}[\frac{3\delta_{ki(j)}+1}{M_{\Phi^{--}}^{2}}
+\frac{1}{M_{\Phi^{-}}^{2}}]^{2}Br(l_{i}\rightarrow
e\nu_{e}\overline{\nu}_{i}).
\end{equation}
Where $\alpha_{e}$ is the fine structure constant and $G_{F}$ is the
Fermi constant. The factor $3\delta_{ki(j)}$ means that, when the
internal lepton is the same as one of the leptons $l_{i}$ and
$l_{j}$, the contributions of $\Phi^{--}$ to
$Br(l_{i}^{-}\rightarrow l_{j}^{-}\gamma)$ is four times those for
$k \neq i$ and $j$. $M_{\Phi^{--}}$ and $M_{\Phi^{-}}$ are the
masses of the scalars $\Phi^{--}$ and $\Phi^{-}$, respectively. In
the $LH$ model, the scalar mass is generated through the
Coleman-Weinberg mechanism and the scalars $\Phi^{--}$, $\Phi^{-}$
and $\Phi^{0}$ degenerate at the lowest order[5]. Thus, we can
assume $M_{\Phi^{--}}= M_{\Phi^{-}}$ and write the branching ratio
as:
\begin{equation}
Br(l_{i}^{-}\rightarrow l_{j}^{-}\gamma)=\frac{\alpha_{e}}{96\pi
G_{F}^{2}M_{\Phi}^{4}}\sum_{k=\tau,\mu,e}(Y_{ik}Y_{kj}^{\ast})^{2}
[3\delta_{ki(j)}+2]^{2}Br(l_{i}\rightarrow
e\nu_{e}\overline{\nu}_{i}).
\end{equation}
In particular, for the decay process $\mu^{-}\rightarrow
e^{-}\gamma$, we obtain the following expression for the branching
ratio $Br(\mu^{-}\rightarrow e^{-}\gamma)$:
\begin{equation}
Br(\mu^{-}\rightarrow e^{-}\gamma)=\frac{\alpha_{e}}{96\pi
G_{F}^{2}M_{\Phi}^{4}}[25(Y_{\mu
e}Y_{ee}^{\ast})^{2}+25(Y_{\mu\mu}Y_{\mu e}^{\ast})^{2}+4(Y_{\mu
\tau}Y_{\tau e}^{\ast})^{2}].
\end{equation}

\begin{figure}[htb] \vspace{0cm}
\begin{center}
\epsfig{file=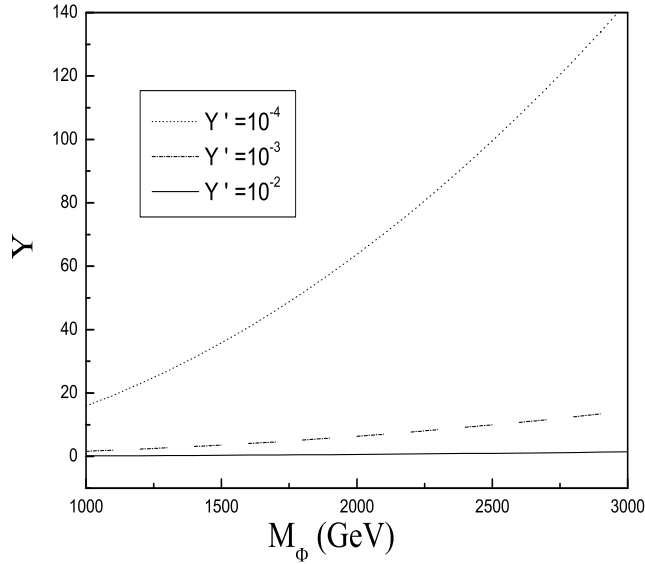,width=270pt,height=250pt} \vspace{-1.0cm}
 \caption{The $FD$ coupling constant $Y$ as a function of the scalar mass
 $M_{\Phi}$ for different \hspace*{1.8cm}values of the $FX$ coupling constant $Y'$.}
\label{ee}
\end{center}
\end{figure}

From above equations, we can see that the $LFV$ process
$l_{i}\rightarrow l_{j}\gamma$ can not be able to constrain $Y_{ij}$
independently. However, if we assume $Y_{ik}=Y$ for $i=k$ ($Y$ is
the flavor-diagonal $(FD)$ coupling constant) and $Y_{ik}=Y'$ for
$i\neq k$ ($Y'$ is the flavor-mixing $(FX)$ coupling constant), then
we can obtain the constraints on the combination of the free
parameters $Y$, $Y'$ and $M_{\Phi}$. Observably, the most stringent
constraint should come from the current experimental upper limits on
the branching ratio $Br(\mu\rightarrow e\gamma)$. Thus, in Fig.2, we
have shown the $FD$ coupling constant $Y$ as a function of the mass
parameter $M_{\Phi}$ for $Y'=1\times 10^{-2}$, $1\times 10^{-3}$ and
$1\times 10^{-4}$. From Fig.2, one can see the upper limit on $Y$
strongly depend on the values of $M_{\Phi}$ and $Y'$. For
$M_{\Phi}\leq 2000 GeV$ and $Y'\geq 1\times 10^{-4}$, there must be
$Y\leq 64$.

In the $LH$ model, the $LFV$ processes $l_{i}\rightarrow
l_{j}l_{k}l_{k}$ can be generated at tree level through the exchange
of doubly charged scalar $\Phi^{\pm\pm}$, as depicted in Fig.3.

\begin{figure}[htb] \vspace{-8.0cm}
\begin{center}
\epsfig{file=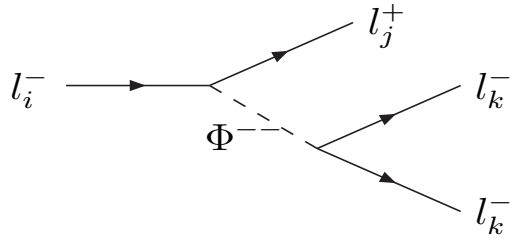,width=800pt,height=1000pt} \vspace{-25.0cm}
 \caption{Tree level Feynman diagram for the $LFV$ processes $l_{i}^{-}\rightarrow
 l_{j}^{+}l_{k}^{-}l_{k}^{-}$ mediated by \hspace*{1.8cm}the doubly charged scalar $\Phi^{--}$.}
 \label{ee} \vspace{-0.5cm}
\end{center}
\end{figure}

The expressions of the branching ratios for the processes
$l_{i}^{-}\rightarrow l_{j}^{+}l_{k}^{-}l_{k}^{-}$ are given
by[16,17]
\begin{eqnarray}
Br(\mu^{-}\rightarrow e^{+}e^{-}e^{-})&=&\frac{\mid Y_{\mu
e}Y_{ee}^{\ast}\mid^{2}}{16G_{F}^{2}M_{\Phi}^{4}},\\
Br(\tau^{-}\rightarrow e^{+}e^{-}e^{-})&=&\frac{\mid Y_{\tau
e}Y_{ee}^{\ast}\mid^{2}}{16G_{F}^{2}M_{\Phi}^{4}}Br(\tau\rightarrow
e\nu_{e}\overline{\nu}_{\tau}),\\
Br(\tau^{-}\rightarrow \mu^{+}e^{-}e^{-})&=&\frac{\mid
Y_{\tau\mu}Y_{ee}^{\ast}\mid^{2}}{32G_{F}^{2}M_{\Phi}^{4}}Br(\tau\rightarrow
e\nu_{e}\overline{\nu}_{\tau}),\\
Br(\tau^{-}\rightarrow e^{+}\mu^{-}\mu^{-})&=&\frac{\mid Y_{\tau
e}Y_{\mu\mu}^{\ast}\mid^{2}}{32G_{F}^{2}M_{\Phi}^{4}}Br(\tau\rightarrow
e\nu_{e}\overline{\nu}_{\tau}),\\
Br(\tau^{-}\rightarrow \mu^{+}\mu^{-}\mu^{-})&=&\frac{\mid
Y_{\tau\mu}Y_{\mu\mu}^{\ast}\mid^{2}}{16G_{F}^{2}M_{\Phi}^{4}}Br(\tau\rightarrow
e\nu_{e}\overline{\nu}_{\tau}).
\end{eqnarray}

Certainly, up to one loop, the $LFV$ processes $l_{i}\rightarrow
l_{j}l_{k}l_{k}$ get additional contributions from the processes
$l_{i}\rightarrow l_{j}\gamma^{\ast}\rightarrow l_{j}l_{k}l_{k}$.
Thus, the charged scalars $\Phi^{\pm\pm}$ and $\Phi^{\pm}$ have
contributions to the $LFV$ processes $l_{i}\rightarrow
l_{j}l_{k}l_{k}$ at one loop. However, compared with the tree level
contributions, they are very small, which can be safely neglected.

The $LFV$ processes $l_{i}\rightarrow l_{j}l_{k}l_{k}$ also can not
give the constraints on the coupling constants $Y_{ij}$
independently, but would be able to constrain the combination $\mid
Y_{ij}Y_{kk}^{\dag}\mid^{2}/M_{\Phi}^{4}$. Our numerical results are
given in Table 1.

In the following section, we will take into account these
constraints coming from the $LFV$ processes $l_{i}\rightarrow
l_{j}\gamma$ and $l_{i}\rightarrow l_{j}l_{k}l_{k}$, estimate the
contributions of the doubly charged scalars $\Phi^{\pm\pm}$ to the
processes $e^{\pm}e^{\pm}\rightarrow l_{i}^{\pm}l_{j}^{\pm}$ and
$e^{+}e^{-}\rightarrow l_{i}^{\pm}l_{j}^{\pm}$, and discuss the
possibility of detecting the signals for the doubly charged scalars
$\Phi^{\pm\pm}$ at the $ILC$ experiments.

\section*{IV. The doubly charged scalars $\Phi^{\pm\pm}$ and the $LFV$
\\ \hspace*{1.2cm}processes  $e^{\pm}e^{\pm}\rightarrow l_{i}^{\pm}l_{j}^{\pm}$
and $e^{+}e^{-}\rightarrow l_{i}^{\pm}l_{j}^{\pm}$}

\begin{figure}[htb] \vspace{-8.0cm}
\begin{center}
\hspace*{-2.5cm} \epsfig{file=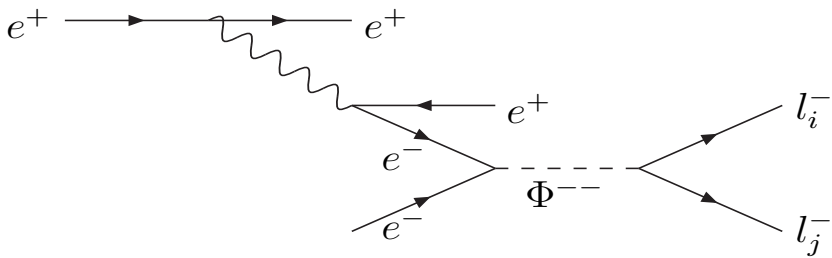,width=800pt,height=1000pt}
\vspace{-24.2cm}
 \caption{Main Feynman diagram for the processes $e^{+}e^{-}\rightarrow l_{i}^{-}l_{j}^{-}$
  predicted by $\Phi^{--}$. }
 \label{ee} \vspace{-0.5cm}
\end{center}
\end{figure}

In general, the doubly charged scalars can not couple to quarks and
their couplings to leptons break the lepton number by two units,
leading to a distinct signature, namely a pair of same sign leptons.
The discovery of a doubly charged scalar would have important
implications for our understanding of the Higgs sector and more
importantly, for what lies beyond the $SM$. This fact has made one
give more elaborate theoretical calculations in the framework of
some specific models beyond the $SM$ and see whether the signatures
of this kind of new particles can be detected in the future high
energy experiments. For example, the production and decay of the
doubly charged scalars and their possible signals at the $ILC$ have
been extensively studied in Refs.[18,19]. In this section, we will
consider the contributions of the doubly charged scalars
$\Phi^{\pm\pm}$ predicted by the $LH$ model to the processes
$e^{\pm}e^{\pm}\rightarrow l_{i}^{\pm}l_{j}^{\pm}$ and
$e^{+}e^{-}\rightarrow l_{i}^{\pm}l_{j}^{\pm}$($l_{i}$ or
$l_{j}$$\neq e$). The processes $e^{\pm}e^{\pm}\rightarrow
l_{i}^{\pm}l_{j}^{\pm}$ can be seen as the subprocesses of the
processes $e^{+}e^{-}\rightarrow l_{i}^{\pm}l_{j}^{\pm}$. For
example, the doubly charged scalar $\Phi^{--}$ generates contributes
to the process $e^{+}e^{-}\rightarrow l_{i}^{-}l_{j}^{-}$ through
the subprocess $e^{-}e^{-}\rightarrow l_{i}^{-}l_{j}^{-}$, as shown
in Fig.4.

Using Eq.(4), the expression of the cross section for the subprocess
$e^{-}e^{-}\rightarrow l_{i}^{-}l_{j}^{-}$ can be easily written as:
\begin{equation}
\widehat{\sigma}(\widehat{s})=\frac{Y_{ee}^{2}Y_{ij}^{2}}{8\pi}\frac{\widehat{s}}{(\widehat{s}-M_{\Phi}^{2})^{2}
+M_{\Phi}^{2}\Gamma_{\Phi}^{2}}.
\end{equation}
Where $\sqrt{\widehat{s}}$ is the center-of-mass $(C.M.)$ energy of
the subprocess $e^{-}e^{-}\rightarrow l_{i}^{-}l_{j}^{-}$.
$\Gamma_{\Phi}$ is the total decay width of the doubly charged
scalar $\Phi^{--}$, which has been given by Ref.[5] in the case of
the triplet scalars $(\Phi^{\pm\pm}, \Phi^{\pm},$ and $\Phi^{0})$
degenerating at lowest order with a common mass $M_{\Phi}$:
\begin{eqnarray}
\Gamma_{\Phi}&=&\sum_{ij}\Gamma(\Phi^{--}\rightarrow
l_{i}^{-}l_{j}^{-})+\Gamma(\Phi^{--}\rightarrow
W_{L}^{-}W_{L}^{-})+\Gamma(\Phi^{--}\rightarrow
W_{T}^{-}W_{T}^{-})\nonumber\\
&\approx&\frac{M_{\Phi}}{8\pi}[3Y^{2}+6Y'^{2}]+\frac{\nu'^{2}M_{\Phi}^{3}}{2\pi\nu^{4}}
+\frac{g^{4}\nu'^{2}}{4\pi M_{\Phi}}.
\end{eqnarray}
Where $Y=Y_{ij}$ $(i=j)$ is the $FD$ coupling constant, $Y'=Y_{ij}$
$(i\neq j)$ is the $FX$ coupling constant. In above equation, the
final-state masses have been neglected compared to the mass
parameter $M_{\Phi}$. It has been shown that, for $\nu'< 1\times
10^{-5}$, the main decay modes of $\Phi^{--}$ are
$l_{i}^{-}l_{j}^{-}$. Furthermore, the FX coupling constant $Y'$ are
subject to very stringent bounds from the $LFV$ process
$\mu\rightarrow eee$. In this case, the decay width $\Gamma_{\Phi}$
can be approximately written as:
\begin{equation}
\Gamma_{\Phi}\approx\frac{3M_{\Phi}Y^{2}}{8\pi}.
\end{equation}
Considering the current bounds on the neutrino mass[8], there should
be:
\begin{equation}
Y_{ij}\nu'\sim10^{-10}GeV,
\end{equation}
so $\nu'< 1\times 10^{-5}$ leads to $Y_{ij}> 1\times 10^{-5}$, which
does not conflict with the most stringent constraint from the $LFV$
process $\mu\rightarrow eee$. Thus, in our numerical calculation, we
will take Eq.(15) as the total decay width of $\Phi^{--}$.

Using the equivalent particle approximation method[20], the
effective cross section for the process $e^{+}e^{-}\rightarrow
l_{i}^{-}l_{j}^{-}$ can be approximately written as[19]:
\begin{equation}
\sigma(E_{e^{+}},s)=\int_{x_{min}}^{1}dx
F_{e^{+}}^{e^{-}}(x,E_{e^{+}}) \widehat{\sigma}(\widehat{s}).
\end{equation}
Where $\widehat{s}=xs$ and $x_{min}=(m_{l_{i}}+m_{l_{j}})^{2}/s$.
$F_{e^{+}}^{e^{-}}(x,E_{e^{+}})$ is the equivalent electron
distribution function of the initial positron, which gives the
probability that an electron with energy $E_{e^{-}}=xE_{e^{+}}$ is
emitted from a positron beam with energy $E_{e^{+}}$. The relevant
expression can be written as[21]:
\begin{equation}
F_{e^{+}}^{e^{-}}(x,E_{e^{+}})=\frac{\alpha_{e}^{2}}{8\pi^{2}x}[ln(\frac{E_{e^{+}}}{m_{e}})^{2}-1]^{2}
[\frac{4}{3}+x-x^{2}-\frac{4}{3}x^{3}+2x(1+x)lnx].
\end{equation}

\begin{figure}[htb]
\begin{center}

\hspace{10cm}\vspace{0.5cm}
\epsfig{file=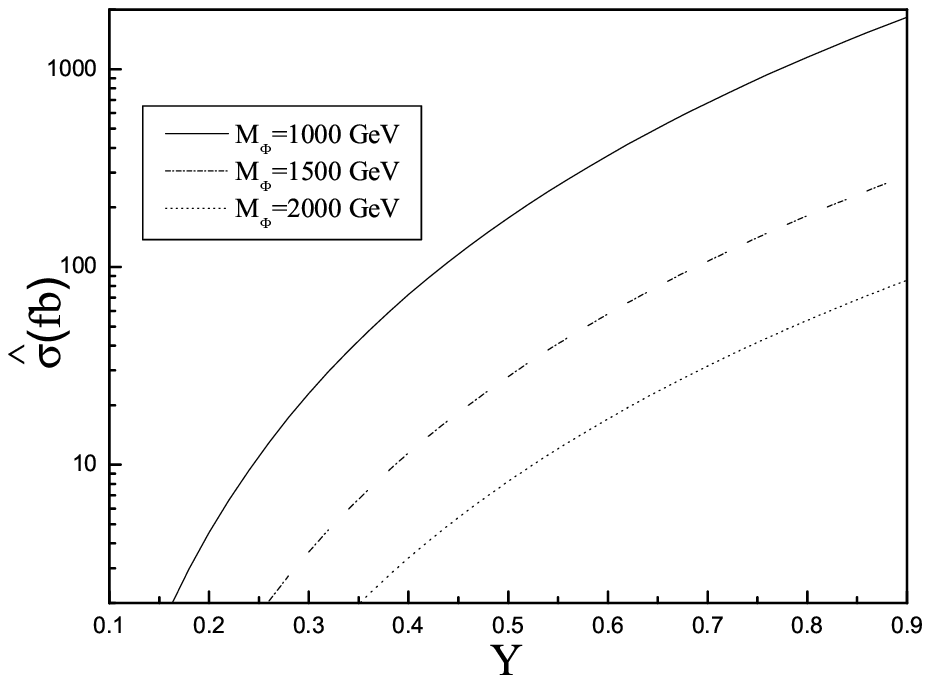,width=220pt,height=200pt}\hspace{-0.5cm}\hspace{0cm}
\vspace{-0.25cm} \put(-230,-8){Figure 5: The cross section
$\widehat{\sigma}(\widehat{s})$ as a function \put(-190,-20){of $Y$
for three values of the mass $M_{\Phi}$.}}
\epsfig{file=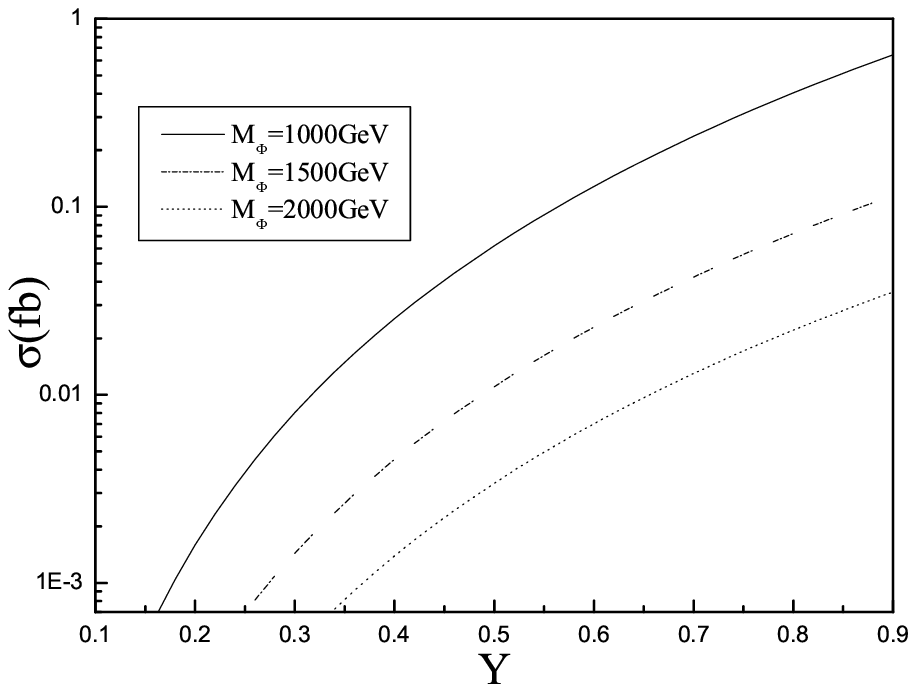,width=220pt,height=200pt} \put(-190,-8){Figure
6: Same as Fig.5 but for $\sigma(s)$. }\hspace{-0.5cm}
 \hspace{10cm}\vspace{-1cm}

\label{ee}
\end{center}
\end{figure}
\vspace{5cm}

\vspace{-4.5cm}In Fig.5 and Fig.6, we plot the production cross
sections $\widehat{\sigma}(\widehat{s})$ and $\sigma(s)$ for the
processes $e^{-}e^{-}\rightarrow \mu^{-}\mu^{-}$ and
$e^{+}e^{-}\rightarrow \mu^{-}\mu^{-}$ as function of the FD
coupling constant $Y$, respectively. In these figures, we have
assumed $0.15\leq Y\leq 0.9$ and taken $\sqrt{s}=500 GeV$ and
$M_{\Phi}=1.0 TeV,1.5 TeV,2.0TeV$. From Fig.5 and Fig.6 one can see
that the values of $\widehat{\sigma}(\widehat{s})$ and $\sigma(s)$
are strongly depend on the value of the $FD$ coupling constant
$Y(Y_{ee})$. For $Y\geq0.7$ and $M_{\Phi}\leq1.5 TeV$, the values of
the subprocess cross section $\widehat{\sigma}(\widehat{s})$ and the
effective cross section $\sigma(s)$ are larger than $1.1\times
10^{2}$ fb and $4.3\times 10^{-2}$ fb, respectively.

The signal of the doubly charged scalar $\Phi^{--}$ given by the
process $e^{+}e^{-}\rightarrow \mu^{-}\mu^{-}$ is so distinctive and
is the $SM$ background free, discovery would be signalled by even
few events. In Fig.7, we plot the discovery region in the
$Y-M_{\Phi}$ plane at 95\% confidence level $(C.L.)$ for seeing 5
$\mu^{-}\mu^{-}$ events, in which we have assumed the future $ILC$
with the $C.M.$ energy $\sqrt{s}=500 GeV$ and the yearly integrated
luminosity of $\mathscr{L}=500fb^{-1}$[22]. From this figure, one
can see that, in wide range of the parameter space, the signals of
$\Phi^{--}$ should be detected in the future ILC experiments.

\begin{figure}[htb] \vspace{0cm}
\begin{center}
\epsfig{file=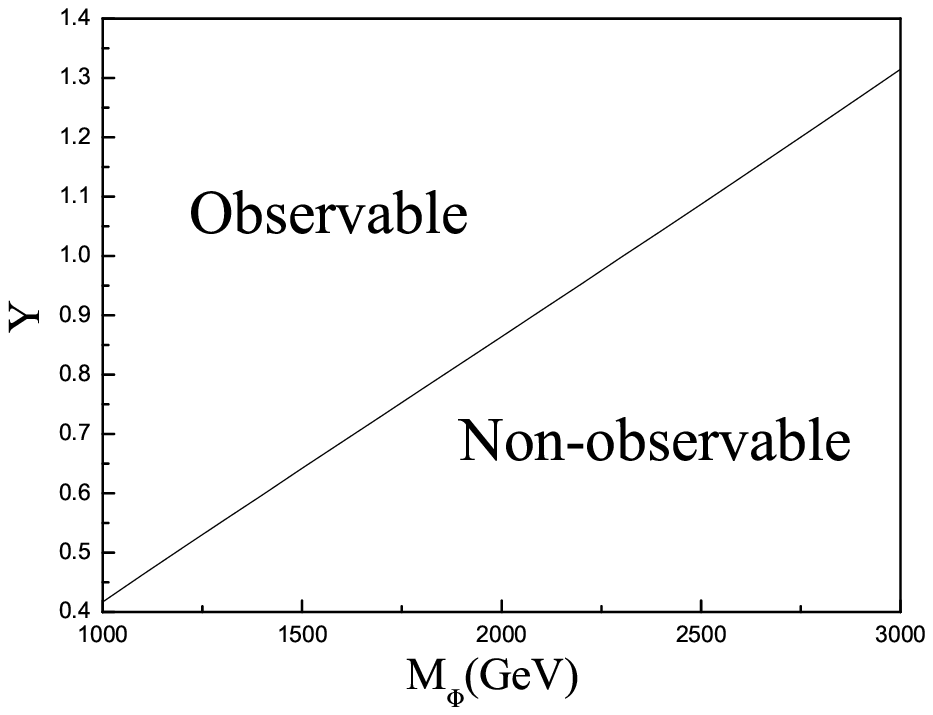,width=270pt,height=250pt} \vspace{-1.0cm}
\put(-360,-5){Figure 7: Discovery region in the $Y-M_{\Phi}$ plane
at 95\% $C.L.$
 for seeing 5 $\mu^{-}\mu^{-}$ events.}
\label{ee}
\end{center}
\end{figure}

\vspace{0.5cm}The doubly charged scalar $\Phi^{--}$ can also has
contributions to the $LFV$ processes
$e^{+}e^{-}\rightarrow\tau^{-}\mu^{-}, \tau^{-}e^{-},$ and
$\mu^{-}e^{-}$. However, the experimental upper limits on the $LFV$
processes $\tau\rightarrow \mu ee$, $\tau\rightarrow eee$, and
$\mu\rightarrow eee$ can give severe constraints on the combination
$\mid Y_{ij}Y_{kk}^{\dag}\mid^{2}/M_{\Phi}^{4}$, which makes the
production cross sections of these processes very small. For
example, even if we take $Y=1$ and $M_{\Phi}\leq 2 TeV$, the
production cross sections $\sigma(\tau\mu),$  $\sigma(\tau e)$, and
 $\sigma(\mu e)$ are smaller than $6.9\times 10^{-3}$fb, $2.1\times
10^{-3}$fb, and $1.9\times 10^{-9}$fb, respectively. Thus, it is
very difficult to detect the signals of $\Phi^{--}$  via the
processes $e^{+}e^{-}\rightarrow l_{i}^{-}l_{j}^{-}(i\neq j)$ in the
future $ILC$ experiments.

Certainly, the doubly charged scalar $\Phi^{++}$ has contributions
to the processes $e^{+}e^{+}\rightarrow l_{i}^{+}l_{j}^{+}$ and
$e^{+}e^{-}\rightarrow l_{i}^{+}l_{j}^{+}$. Similar with above
calculation, we can give the values of the production cross sections
for these processes. We find that the cross section
$\sigma(l_{i}^{+}l_{j}^{+})$ is equal to the cross section
$\sigma(l_{i}^{-}l_{j}^{-})$. Thus, the conclusions for the doubly
charged scalar $\Phi^{--}$ are also apply to the doubly charged
scalar $\Phi^{++}$.

\section*{V. Conclusions }

\hspace{5mm}To solve the so-called hierarchy or fine tuning problem
of the $SM$, the little Higgs theory was proposed as a kind of
models to $EWSB$ accomplished by a naturally light Higgs boson. The
$LH$ model is one of the simplest and phenomenologically viable
models. In the $LH$ model, neutrino masses and mixings can be
generated by coupling the scalar triplet $\Phi$ to the leptons in a
$\bigtriangleup L=2$ interaction, whose magnitude is proportional to
the triplet $VEV \nu'$ multiplied by the Yukawa coupling constant
$Y_{ij}$ without invoking a right handed neutrino. This scenario
predicts the existence of doubly charged scalars $\Phi^{\pm\pm}$.
For smaller values of $\nu'$ i.e. $\nu'\leq 1\times 10^{-5}$, the
doubly charged scalars $\Phi^{\pm\pm}$ have large flavor changing
coupling to leptons, which can generate significantly contributions
to some $LFV$ processes and give characteristic signatures in the
future high energy experiments.

In this paper, we first consider the $LFV$ processes
$l_{i}\rightarrow l_{j}\gamma$ and $l_{i}\rightarrow
l_{j}l_{k}l_{k}$ in the context of the $LH$ model. For the $LFV$
process $l_{i}\rightarrow l_{j}\gamma$, it involves all of the FX
coupling constants $Y_{ij} (i\neq j)$, we can not give the simple
constraints about the free parameters $Y_{ij}$ and $M_{\Phi}$. Thus,
for the fixed values of the FX coupling constant $Y'=Y_{ij}(i\neq
j)$, we take into account the current experimental upper limit of
the $LFV$ $\mu\rightarrow e\gamma$ and plot the FD coupling constant
$Y=Y_{ij}(i=j)$ as a function of the mass parameter $M_{\Phi}$. Our
numerical results show that the upper limit on $Y$ is strongly
depend on the free parameters $M_{\Phi}$ and $Y'$.

Using the present experimental upper limits on the branching ratios
$Br(l_{i}\rightarrow l_{j}l_{k}l_{k})$, we obtain the constraints on
the combination $\mid Y_{ij}Y^{*}_{kk}\mid^{2}/M_{\Phi}^{4}$. We
find that the most stringent constraint comes from the $LFV$ process
$\mu\rightarrow eee$. In all of the parameter space, there must be
$\mid Y_{\mu e}Y_{ee}^{\ast}\mid^{2}/M_{\Phi}^{4}\leq 2.2\times
10^{-19} GeV^{-4}$.

The characteristic signals of the processes $e^{+}e^{-}\rightarrow
l_{i}^{\pm}l_{j}^{\pm}$ is same-sign dileptons or two same-sign
different flavor leptons, which is the $SM$ background free and
offers excellent potential for doubly charged scalar discovery. To
see whether the doubly charged scalar $\Phi^{--}$ can be detected in
the future $ILC$ experiments, we discuss the contributions of
$\Phi^{--}$ to the processes $e^{-}e^{-}\rightarrow
l_{i}^{-}l_{j}^{-}$ and $e^{+}e^{-}\rightarrow l_{i}^{-}l_{j}^{-}$.
We find that the triplet scalar $\Phi^{--}$ can give significantly
contributions to the processes $e^{+}e^{-}\rightarrow
l_{i}^{-}l_{i}^{-}$. In wide range of the parameter space of the
$LH$ model, the possible signals of $\Phi^{--}$ might be observed in
the future $ILC$ experiments. However, the production cross sections
of the $LFV$ processes $e^{+}e^{-}\rightarrow
l_{i}^{-}l_{j}^{-}(i\neq j)$ mediated by $\Phi^{--}$ are very small.
The contributions of the triplet scalar $\Phi^{++}$ to the processes
$e^{+}e^{-}\rightarrow l_{i}^{+}l_{j}^{+}$ are equal to those of
$\Phi^{--}$ for the processes $e^{+}e^{-}\rightarrow
l_{i}^{-}l_{j}^{-}$, Thus, our conclusions are also apply to the
doubly charged scalar $\Phi^{++}$.

Some popular models beyond the $SM$ predict the existence of doubly
charged scalars, which generally have the lepton number and lepton
flavor changing couplings to leptons and might produce distinct
experimental signatures in the current or future high energy
experiments. Their observation would signal physics outside the
current paradigm and further test the new physics models. Search for
this kind of new particles has been one of the important goals of
the high energy experiments[23]. Thus, the possibly signals of the
doubly charged scalars $\Phi^{\pm\pm}$ predicted by the little Higgs
models should be more studied in the future.

\vspace{0.5cm}

\noindent{\bf Acknowledgments}

This work was supported in part by Program for New Century Excellent
Talents in University(NCET-04-0290), the National Natural Science
Foundation of China under the Grants No.10475037 and 10675057.

\null
\end{document}